\documentstyle[preprint,aps,eqsecnum]{revtex}
\begin{document}
\preprint{UCD 94-21, July 1994}
\title{Phase of the Wilson line}
\author{Joe Kiskis\thanks{email: jekucd@ucdhep.ucdavis.edu  }}
\address{
Department of Physics\\
University of California, Davis, CA 95616, USA}
\date{\today}
\maketitle
\begin{abstract}
This paper discusses some aspects of the global $Z(N)$ symmetry of
finite-temperature, $SU(N)$, pure Yang-Mills lattice gauge theory.
It contributes to the recent discussion of the physics of the
phase of the Wilson line expectation value.
In the high $T$ phase, $\langle L \rangle$ takes one of
$N$ distinct values proportional to the $Nth$ roots of unity in $Z(N)$, and
the $Z(N)$ symmetry is broken.
Only one of these is consistent with the usual interpretation
$\langle L \rangle = e^{-F/T}$ with $F$
the excess free energy due to a source of color-electric flux at the position
of the line.
This relation should be generalized to $\langle L \rangle=ze^{-F/T}$ with
$z \in Z(N)$ so that it is consistent with the
negative or complex values.
In the
Hamiltonian description,
the {\em physical} variables are the group elements on the links of the spatial
lattice.
In a Lagrangian formulation, there are also group
elements on links in the inverse-temperature direction from which the Wilson
line
is constructed. These are unphysical, auxiliary variables introduced
to enforce the Gauss law constraints.
The following results are obtained: The generalized relation
$\langle L \rangle=ze^{-F/T}$, which has appeared in earlier papers,
is derived.
The value of $z \in Z(N)$ is determined by the external field that is needed
for taking
the infinite-volume limit. There is a single
physical, high-temperature phase, which is the same for all $z$.
The global $Z(N)$ symmetry is not physical; it acts as the identity on all
physical states. In the Hamiltonian formulation, the high-temperature
phase is not distinguished by physical broken symmetry.
Rather the high-temperature
phase has a percolating flux network that is not present in the low-temperature
phase.
\end{abstract}
\pacs{11.15.Ha, 5.50.+q, 75.10.Hk, 64.60.Ak}
\narrowtext

\section{Introduction}

This paper discusses some aspects of the global $Z(N)$ symmetry\cite{s1} of
finite-temperature, $SU(N)$, lattice gauge theory without matter
fields. There
is a low-temperature phase, which is presumed to confine fundamental sources,
and a high-temperature phase, which does not.
The order parameter is the
finite-temperature expectation value of the Wilson line $\langle L \rangle$.
The fundamental-representation
line $L(i)$ is the normalized trace of the
product of the group elements on the links in the inverse-temperature direction
at spatial site $i$.
\begin{equation}
 \langle L \rangle = \langle L(i) \rangle = \langle \case{1}{N}
                       Tr[\prod_{i_4} U(i,i_4,4)]\rangle
\end{equation}
The line carries a nontrivial representation of the
$Z(N)$ symmetry. In the confining phase, $\langle L \rangle=0$, and the
ensemble
is $Z(N)$ symmetric. In the high $T$ phase, $\langle L \rangle$ takes one of
$N$ distinct values proportional to the $Nth$ roots of unity in $Z(N)$.
The $Z(N)$ symmetry is broken.
A combination of these pure phases would not have cluster decomposition
for the Wilson line n-point functions.
The broken-symmetry, pure phases appear in Monte Carlo simulations, where runs
of finite duration in the
high-temperature phase and not too close to the critical temperature show the
Wilson line fluctuating around one of the $N$ available values.

In a Hamiltonian approach\cite{s2}, the line is a projection operator that
forces the
gauge field to be in a fundamental rather than a singlet state at the spatial
position of the line. Thus it is usually interpreted as the
exponential of the excess free energy due to the source: $\langle L
\rangle=e^{-F/T}$. This implies that $\langle L \rangle$
should be positive. Thus, there is a problem. The physical
interpretation should be refined so that is consistent with negative
or complex values for $\langle L \rangle$ in the
nonconfining, high-temperature phase.

There is discussion of this point in recent papers\cite{r2}.
Appearing there, in various contexts, is
a modified relation between the line and the
excess free energy $\langle L \rangle=ze^{-F/T}$ with $z \in Z(N)$. There is
also discussion of the possibility that the $Z(N)$ symmetry is not physical and
that the $N$ high-temperature, pure phases associated with the $N$ possible
values for $\langle L \rangle$ are not physically distinct. As a consequence,
well-defined interfaces may not exist.

Here, I offer a discussion of these issues that emphasizes simplicity and gives
a short, direct connection between basic principles and results. I use the
Hamiltonian description of finite-temperature gauge theory on a lattice in
three spatial dimensions in the $A_0=0$ gauge with Gauss law constraints.  In
this formulation, the physical interpretation is most accessible, and the
results come fairly easily. I will refer to this as the {\em physical} theory.
The {\em physical variables} are the group elements on the links of the spatial
lattice. The constraints are enforced through the use of additional group
elements on sites. These are unphysical, auxiliary variables.
In a four-dimensional, Lagrangian formulation, they become the
elements on links in the inverse-temperature, 4-direction. The Wilson line
is constructed from them.
I obtain the following results: The relation
$\langle L \rangle=ze^{-F/T}$ is verified.
The value of $z$ is determined by the external field that is needed for taking
the infinite-volume limit in the high-temperature phase.  There is a single
physical, high-temperature phase, which is independent of the value of $z$.
The global $Z(N)$ transformation is not physical: it acts as the identity on
all
physical states. Thus, in the Hamiltonian formulation, the high-temperature
phase is not distinguished by broken symmetry. Rather the high-temperature
phase has a percolating flux network that is not present in the low-temperature
phase, where there are only finite clusters of flux.

There is considerable overlap of the present paper with Refs.\ \cite{r2}.
Some of the statements in those papers are confirmed or derived by other
means here. There are several additional points that I emphasize: It is
important to use an external field and some care in the infinite-volume limit.
The value of the factor $z$ in $\langle L \rangle$ is determined by that
external field. Even in a pure
phase with $\langle L \rangle \neq 0$, the local projection operators function
correctly to enforce the Gauss law constraints. It is percolating flux rather
than any {\em physical} symmetry breaking that distinguishes the nonconfining
phase from the confining one.

These points are discussed first in the context of a simple
model for flux on links. It can be interpreted
as a degenerate $Z(2)$ gauge theory and is equivalent to the three-dimensional
Ising model.

With this Ising flux model understood,
it is easy to repeat the analysis for $SU(N)$ gauge theory.

In Sec.\ II, the Ising flux model is discussed. Section III has the treatment
of
$SU(N)$ gauge theory. Section IV discusses percolating flux.
There is a brief conclusion in Sec.\ V.

\section{Ising flux}

This section discusses a model which is simpler than
$SU(N)$ gauge theory. However, it contains the essential elements of the
problem at hand. It is a model for flux on links that is motivated by
finite-temperature, $SU(2)$, gauge theory and is equivalent to the Ising model
in three dimensions. It can also be viewed as a $Z(2)$ gauge theory
without spatial plaquettes in the action.
The analogue of Wilson lines on sites is spins on sites so
$\langle L \rangle \rightarrow \langle s \rangle$.
Some of the critical properties of this model are discussed in Ref.\ \cite{r3}.
Here the model serves as a convenient arena in which to introduce the approach
that is used on $SU(N)$ gauge theory in Sec.\ III.

For the Ising flux model, there are results that are analogous to those for
full
gauge theories:
In the high-temperature phase, $\langle s \rangle=\pm e^{-F/T}$
with the sign determined by
the external field. There is a single physical, high-temperature phase,
which is independent of the sign of $\langle s \rangle$. The global $Z(2)$
transformation acts trivially on physical states.
The high-temperature phase is
not distinguished by broken symmetry. Rather the high and low-temperature
phases differ in that the high-temperature phase has a percolating flux network
whereas the low-temperature phase has only finite clusters of flux. This last
point was covered in some detail previously\cite{r3}.

\subsection{Flux model details}

In the Hamiltonian, strong-coupling description of $SU(2)$ lattice gauge
theory,
the flux on links is labeled by the irreducible representations of $SU(2)$. At
each site, the representations must combine to have a singlet piece.  If all of
the nontrivial flux is in the fundamental representation of $SU(2)$, then there
must be an even number of links with flux at each site.

The flux model results from further simplifications. First reduce the gauge
group from $SU(2)$ to its center $Z(2)$. Consider it in a Hamiltonian or
transfer matrix formalism. Then delete the contributions to the Hamiltonian
from the spatial plaquettes. With that, the Hamiltonian is very easily
diagonalized. All that remains is the energy of the links that are in the
nontrivial rather than the trivial representation of $Z(2)$. Those are the
links with flux.

In this simplified model, space is a three-dimensional cubic lattice.
There are variables $\theta$ on links that can have the values 1 or 0 to
indicate the presence or absence of flux.
These are the physical variables.
A configuration is specified by the function $\theta(l)$. States with definite
flux $|\theta\rangle$ are labeled by that function.
A general state has a wave
functional that gives the amplitude for the system to be found in the various
basis states of definite flux.
The energy of a link with flux is
$\sigma$. The weight for a configuration is
\begin{equation}
  \langle \theta | e^{-H/T} | \theta \rangle =
 e^{- \case{1}{T} \sum_{l} \sigma \theta(l)} =
            \prod_{l} e^{- \case{1}{T} \sigma \theta(l)}   . \label{e1}
\end{equation}
The sum over configurations is restricted to those in which the number of links
at a site that have flux is even. Let the collection of all such
configurations be $C'$. It is a subset of the unrestricted collection
of all configurations $C$. The partition function is
\begin{equation}
 Z = Tr'[e^{-H/T}] \equiv \sum_{C'}
                           \langle \theta | e^{-H/T} | \theta \rangle
      = \sum_{C'} e^{- \case{1}{T} \sum_{l} \sigma \theta(l)} .
\end{equation}

The equivalence of this flux model to the Ising model
results from using site variables to enforce the restriction
on configurations, and then doing the $\theta$ sums\cite{r3}.
The site variables are the Ising spins.
First, consider the sum of $\theta(l)$
over the $2d=6$ $l$'s contained in the set $I(i)$ of links with endpoint $i$:
\begin{equation}
  \Sigma(i) \equiv \sum_{l \in I(i)} \theta(l)  .
\end{equation}
To force this to be even at each site, introduce the site variables $s(i)$,
which take the values $\pm 1$. The factor
\begin{equation}
 \case{1}{2} \sum_{s(i)=\pm 1} s(i)^{\Sigma(i)}            \label{e2}
\end{equation}
has the desired effect. A factor like this is introduced into the
partition function sum for each site $i$. Now the restriction on the $\theta$
sums can be relaxed. With the abbreviations
\begin{equation}
 \sum_s \equiv \prod_i (\case{1}{2} \sum_{s(i)}) ,
\end{equation}
$z=e^{- \sigma/T}$, and $ss'$ for the product of the pair of spins
on the sites that bound a given link, the partition function is
\begin{eqnarray}
 Z & = & \sum_s \sum_{C} \prod_{l} e^{- \case{1}{T}  \sigma \theta(l)}
        \prod_i s(i)^{\Sigma(i)}   \label{e23}  \\
   & = & \sum_s \sum_{C} \prod_{l}
            (z ss')^{\theta(l)}      \label{e4}     \\
   & = & \sum_s \prod_l ( 1 + z ss')   .
\end{eqnarray}
This is a version of the Ising model
\begin{equation}
   Z \sim \sum_s e^{\beta \sum_l ss'}
\end{equation}
with $z=\tanh \beta$.
The links with flux are those that receive the term
$zss'$ rather than 1 in the expansion of $\prod_l ( 1 + z ss')$.
Note that I never refer to the Ising model $1/\beta$ as temperature.
Temperature always refers to the $T$ first appearing in (\ref{e1}).
Small $T$ gives small $z$, small $\beta$, and disordered Ising spins.

{}From the perspective of this paper, the physical variables are flux on
links, and the physical states are the flux configurations.
The Ising spins are unphysical, auxiliary variables
introduced to enforce the restrictions on flux configurations. They are the
analogue of $A_0$ or of the gauge group elements on links in the
inverse-temperature direction.

Now consider a quantum, Hamiltonian view of the constraints.
With $i$ a site of the spatial lattice and $g(i) \in Z(2)$,
let $V(i,g(i))$ be
defined by its action on the basis states of definite flux by
\begin{equation}
    V(i,g(i)) | ... \theta(l)...> =  g(i)^{\Sigma(i)} | ... \theta(l)...>  .
                  \label{e3}
\end{equation}
(Strictly speaking, one should distinguish whether a link contributes $g$ or
$g^{-1}$. However in $Z(2)$, they are equal, and it does not matter.)
Evidently $V$ commutes with the Hamiltonian. It is the realization of a gauge
transformation for this model in the definite-flux basis.
A full gauge transformation is effected by $V(g)=\prod_{i} V(i,g(i))$.
A global $Z(2)$ transformation has $g(i)=\bar{g}$ independent of $i$ and is
written $\bar{V}(\bar{g})$.

The irreducible representation of the group $Z(2)_i$ at site $i$
carried by $|\theta\rangle$  can be the
trivial one corresponding to $\Sigma(i)$ even or the nontrivial one
corresponding to $\Sigma(i)$ odd. The constraint on configurations (analogue of
the Gauss law) says that the representation must be trivial at all $i$.
Using $r=0$ or 1 to indicate the even or odd representations,
respectively,
the projection operator for the representation $r(i)$ at $i$ is
\begin{equation}
 P(i,r(i)) = \case{1}{2}\sum_g  g^{r(i)} V(i,g)   .   \label{e21}
\end{equation}
To select the trivial representation at all sites, use
\begin{equation}
 {\cal P} = \prod_i P(i,0)  .        \label{e22}
\end{equation}
Comparing (\ref{e3}), (\ref{e21}), and (\ref{e22}) with (\ref{e2})
and (\ref{e23}) and identifying $g$
with $s$, shows that the spin sums are the $g$ sums for the projection
operators. Similarly, the numerator of the n-point function
$\langle s(i) s(j) \ldots \rangle$ has projections onto the odd representation
at the sites $i,j,\ldots$ and onto the even representation at all others,
while the denominator has projections onto the trivial representation at all
sites. This gives an interpretation of
\begin{equation}
   \langle s(i) s(j) \ldots \rangle = \frac{Tr[e^{-H/T}P(i,1)P(j,1)\ldots
                                           \prod_{k\neq i,j,\ldots}P(k,0)]}
                                  {Tr[e^{-H/T}\prod_k P(k,0)]}  \label{e5}
\end{equation}
as $e^{-F/T}$ with $F$ the excess free energy due to odd sites at
$i,j,\ldots$.

\subsection{Global ${\bf Z(2)}$ symmetry}

The flux model is equivalent to the familiar three-dimensional
Ising model with spins $s=\pm1$ on sites. There is a global $Z(2)$ symmetry
$s\rightarrow-s$. In the region of broken symmetry $T>T_c$, there are two pure
phases with
$\langle s \rangle$ greater than zero in one and less than zero in the other.
However for the flux model, the spins are
unphysical so that there is not necessarily any
physical symmetry breaking. In fact, there is not.

Consider this in the Hamiltonian view.
The global operators $\bar{V}(g)$ act trivially on all states whether $g=1$ or
$g=-1$. The first is a tautology because $\bar{V}(1)=1$. For the second,
notice that in the global version of (\ref{e3}), each link
appears twice and gives the factor $g(i)^{\theta(l)}g(i')^{\theta(l)}$.
If the group
elements $g(i)$ and $g(i')$ at the two ends of the link $l$ are equal, then
this factor is one. Thus, the global transformations act trivially on the
states.

That means that one can insert the operator $\bar{V}(-1)$ into any matrix
element without changing its value. On the other hand,
\begin{equation}
   \bar{V}(-1) V(g) =\bar{V}(-1) \prod_i V(i,g(i)) = \prod_i V(i,-g(i))  .
\end{equation}
This implies that in $Z$ the weight for the $g$ sums from the projectors is
even in global  $g \rightarrow -g$. This is the same as the Ising model
global symmetry $s \rightarrow -s$.

The crucial point is that $\bar{V}(-1)$ acts as 1 on physical states so that
the global $Z(2)$ transformation of the spins is not a physical transformation.
Thus when the spins order, there is no physical symmetry breaking because there
is no physical symmetry to break.

This can be viewed in another way. Think of doing the $\theta$ sums first with
fixed spins (or $g$'s) and then doing the spin sums. The weights for the
flux configurations depend upon the values of the spins. However, if
all the spins are reversed, there is no effect on the weights of the flux
configurations in (\ref{e4}). When the global $Z(2)$ is broken, there
are two pure phases for the
spins, but the configurations of flux are independent of which is selected.
Thus the physical variables are not affected by the selection of a single
pure, spin phase as is needed for cluster decomposition in spin variables.
This is the analogue of the statement that the gauge theory pure
phases with different values of $\langle L \rangle$ are physically
indistinguishable.
In calculating physical matrix elements, one could
just as well sum over the two pure phases. This would give the same physical
matrix elements but $\langle s \rangle =0$.

So if it is not physical symmetry breaking that distinguishes the
confining and the nonconfining phases, then what is it? It is the presence of a
percolating network of flux. The physical percolating network allows the
nonphysical spins to order. This will be discussed further in Sec.\ IV.

\subsection{Physical interpretations}

Since the physical states of flux are the same for either sign of
$\langle s \rangle$, how should the sign and magnitude of
$\langle s \rangle$ be interpreted?
As shown, the sum on the Ising spin at a site can be
interpreted as projection onto the subspace of states with even flux
at that site. In the same way, the expectation value $\langle s(i) \rangle$
or the correlation function $\langle s(i) s(j) \rangle$ projects onto states
with
odd flux at the indicated sites and even flux elsewhere.
As shown in (\ref{e5}), these objects
are sums of diagonal matrix elements of projection operators.
They are interpreted as $e^{-F/T}$ where
$F$ is the excess free energy of the ensemble with odd flux at some sites.
Thus, these expectation values should be positive.
This is in conflict with the fact
that $\langle s \rangle$ can be less than zero for $T>T_c$.
The apparent contradiction
casts some doubt on the action of the supposed projection operators when the
spins order.
Evidently the problem is related to the infinite-volume limit. In finite
volume, there is no ordering of the spins.

One approach is to finesse the problem by ignoring the single spin expectation
value and looking at the two-point function, which is even in the reflection
operation and does not suffer the problem of interpretation. The two-point
function $\langle s(0) s(i) \rangle$ has the same value in the two
broken-symmetry, pure phases.
The interpretation $\langle s(0) s(i) \rangle =
e^{-F_2(i)/T}$ with $F_2(i)$ the excess
free energy of two odd sites separated by distance $i$ is consistent.
Take $i \rightarrow \infty$ to obtain $F_2(\infty)=2F$ and
$e^{-F/T}=\langle s(0)s(\infty) \rangle^{1/2}=
(\langle s \rangle\langle s \rangle)^{1/2}=|\langle s \rangle|$. This confirms
the relation between $F$ and $|\langle s \rangle|$.
However, it leaves the sign of $\langle s \rangle$ uninterpreted and the
contradiction unresolved.

A technique for handling this kind of problem is to introduce an external
field $h$ and take limits in the order $h \rightarrow 0$ after volume ${\cal V}
\rightarrow \infty$.  The trivial action  of $\bar{V}(-1)$ on physical states
gives the reflection symmetry for the spins. It follows that the free energy is
an even function of $h$. When the volume is finite, there is an expansion in
powers of $h$, $\langle s \rangle \rightarrow 0$ as $h \rightarrow 0$,
and there is no problem. However
for $T>T_c$ and ${\cal V} \rightarrow \infty$, the free energy has a
nonanalyticity of the form
$|h|$. This gives  $\langle s \rangle \sim \epsilon(h)$. It shows that the
sign of $\langle s \rangle$ simply reflects  whether $h \rightarrow 0^+$ or  $h
\rightarrow 0^-$.

It was shown above that the configurations of flux are unaffected by a global
reversal $s \rightarrow -s$. It follows that the weights for flux
configurations are unaffected by $h \rightarrow -h$. Thus whatever $\langle s
\rangle$ has to say about flux, it says the same thing for
$h \rightarrow 0^+$ or $h \rightarrow 0^-$.

So what does $\langle s \rangle$ say about flux? For $h=0$ in finite volume,
$\langle s \rangle = 0$, and the interpretation is very simple: Projection onto
a state with one odd-flux site is zero because it is a geometrical
impossibility to have exactly one odd-flux site.
The $h=0$, infinite-volume limit does not exist. If it did, it would surely
give $\langle s \rangle=0$, which is not correct if spin cluster decomposition
is desired.
With $h \neq 0$, the infinite-volume limit is defined.
The external field gives an
extra factor of  $e^{h \sum_i s(i)}$. To interpret it, write out the
character expansion
\begin{eqnarray}
   e^{h \sum_i s(i)} & = & \prod_i e^{h s(i)} \\
                     & = & \prod_i (\cosh h + s(i) \sinh h ) .
\end{eqnarray}
This shows that the external field inserts odd-flux sites with a density
proportional to $\tanh h$, which vanishes as $h \rightarrow 0$.

In finite volume, there is a small-$h$ expansion that begins
\begin{equation}
  \langle s(0) \rangle_h = \tanh h \sum_i \langle s(0) s(i) \rangle + ...
          \;\;  .
\end{equation}
In this expansion, the n-point functions on the right hand side are evaluated
at $h=0$, and only those with even n appear.
For odd n, the correlation
functions in finite volume with $h=0$ are zero.

The leading, large-volume terms sum to a
function of the form $t({\cal V} h)$.
The function $t(x)$ must have a small $x$ expansion with
odd powers of $x$ and a large $x$ limit with $t(\pm \infty) = \pm
\langle s \rangle$.

This shows that $\langle s \rangle$ is determined by the infinite-volume limit
of n-point functions with n even. These are unaffected by global sign
changes so the sign of $\langle s \rangle$ is entirely determined by the sign
of $h$. Since the flux configurations are the same for either sign of $h$, the
sign of $\langle s \rangle$ is unphysical.

When the spins order, one pure phase with a definite sign for
$\langle s \rangle$ is selected. One does not ordinarily sum over the two pure
phases. Does this mean that some global part of the constraints or projections
is not being
handled correctly? No, because the action of $\bar{V}(-1)$ on physical states
is trivial.

For further peace of mind, one can verify that the local projectors are
functioning properly. This means that there should be
contributions only from configurations with even flux at each site
in a single, ordered, pure phase. A measure of this comes from the
operator
${\cal F}(i)=V(i,-1)=(-1)^{\Sigma(i)}$. If the expectation value of this is
1, then only even-flux
sites are contributing. Inserting this operator is the
same as reversing the spin at site $i$. Thus,
\begin{equation}
 \langle {\cal F}(i) \rangle =  \frac{\sum_{s} [e^{-S}]_{s(i)
                                                       \rightarrow -s(i)}}
                       {\sum_{s} [e^{-S}]}    .
\end{equation}
But a change of summation variable $s(i) \rightarrow -s(i)$ on the top
reduces this ratio to 1 whatever the phase. For more detail, see Sec.\ III.C.

\subsection{Lagrangian view}

For a full analogy with ordinary gauge theory,
this Hamiltonian formulation in three spatial dimensions can be related
to a 4-dimensional Lagrangian formulation in which the lattice has finite
extent and periodic boundary conditions in the inverse-temperature,
4-direction.
Consider a gauge theory with elements of $Z(2)$ on the links of this
4-dimensional lattice. Choose an action with the usual contributions
from the plaquettes with links in the inverse-temperature direction but with
no terms from spatial plaquettes.
Now
choose a gauge where all 4-direction link variables are 1 except those
on a particular
slice at fixed $i_4$. These are the Ising spins.
To obtain the effective theory for these spins, sum over the variables on
spatial links. An Ising interaction among the spins remains.
Thus, the Ising model is a $Z(2)$ gauge theory without terms in the action from
spatial plaquettes.

In this Lagrangian view, there is a global $Z(2)$ symmetry operation that
multiplies all the variables on 4-direction links at a given $i_4$ by $-1$.
Viewed as a classical theory on the 4-dimensional lattice, the
finite-temperature gauge theory has a global $Z(2)$ symmetry.

This is the most common description of finite-temperature gauge theories.
Viewed as classical theories, there is a $Z(2)$ symmetry that can be broken.
But, as discussed above, when they are viewed as quantum systems in three
spatial dimensions at finite temperature, there is no physical $Z(2)$ symmetry.

\subsection{Ising flux conclusion}

For the Ising flux model, whatever the phase, only configurations with even
flux at each site contribute to the sums for the partition function. The
weights of the flux configurations are the same in the two pure phases of
ordered spins. There is no physical difference between these phases.
The expectation value of a spin is
$\langle s \rangle= \epsilon(h) e^{-F/T}$. The quantity $F$ is the excess
free energy for the ensemble
with a source at one point and a background of other sources in the limit of
vanishing density for the other sources. The sign, which is determined by $h$,
is an artifact of the limiting process and has no physical significance.

\section{${\bf SU(N)}$ gauge theory}

Now that the concepts have been developed in an elementary setting,
it is relatively easy to generalize to $SU(N)$ gauge theory.
First there is a short paragraph on the usual Lagrangian description of the
finite-temperature, $Z(N)$ symmetry and its breaking.
Then there is a discussion using the Hamiltonian machinery.
That is followed by a brief mention
of the analogue of the Ising spins view.

The main difference from the Ising flux model
is that states of definite flux are no longer eigenstates of the
Hamiltonian or of the transfer matrix. However, that is not an essential part
of the argument. The important thing is that they are a basis, and that they
still are. In fact, it is only a convenience to work in that basis. The
arguments can be made just as well in the other commonly used basis where the
group elements on links have definite values. The
global transformation by $g=-1$ acts trivially on physical states,
but it produces the global $Z(N)$ symmetry of finite-temperature
gauge theory. It is a symmetry of the unphysical, auxiliary group elements on
links in the 4-direction.  So it is not {\em physical} symmetry breaking that
distinguishes  the confining from the nonconfining phases. This is evident in
either basis. However, the definite-flux basis is clearly more convenient in
describing the percolating flux that does distinguish the confining from the
nonconfining phases.

In the Lagrangian view, finite-temperature gauge theory is a classical
system  of fields on a 4-dimensional lattice with one finite, periodic
direction. The group elements on the 4-direction links have no special
interpretation. There is a global symmetry that results from multiplying all
the group elements on links in the 4-direction at a given $i_4$ by an element
of the center of the group. The breaking of this symmetry, as detected by the
Wilson line, reveals the deconfining phase transition. To give this a
quantum, finite-temperature, physical interpretation, one requires a
Hamiltonian description.

\subsection{Hamiltonian Machinery}

The Hamiltonian description\cite{s2} of the finite-temperature gauge theory on
a
lattice in three spatial dimensions in the $A_0=0$ gauge with Gauss law
constraints is what I refer to as the physical theory. The physical
variables are the group elements on the links of the spatial lattice. The basis
states for the quantum system can be taken as either those with definite flux,
which is definite representation functions on links, or those with definite
group elements.

This state space carries a representation of the gauge group
${\cal G} = \prod_i SU(N)_i$. As before, let $V(i,g(i))$ be the operator for a
gauge transformation by $g(i) \in SU(N)_i$ at site $i$. To enforce the Gauss
law
constraint at site $i$, one projects onto the subspace that carries the trivial
representation (now indicated by {\bf 1}) of this
group action. The operator that does this is
\begin{equation}
  P(i,{\bf 1}) = \int\!\!dg(i)\, V(i,g(i)) .    \label{e51}
\end{equation}
If there are no external sources, then
${\cal P} = \prod_i P(i,{\bf 1})$
projects onto allowed states.
The partition function is
\begin{equation}
  Z = Tr[ e^{-H/T} {\cal P}] .
\end{equation}
When $Z$ is converted to a path integral, the $g(i)$ integrals from (\ref{e51})
become the
integrals over the group elements on the 4-direction links.

To account for an external source at site $i$ in representation $r(i)$, the
projector $P(i,{\bf 1})$ in the product ${\cal P}$ is replaced by one with a
factor of the
character for irreducible representation $r(i)$
\begin{equation}
  P(i,r(i)) = \int\!\!dg(i)\, \chi_{r(i)}^*(g(i))\, V(i,g(i)) .
\end{equation}
In the path integral, this becomes a Wilson line in representation $r(i)$ at
site $i$. In this paper, only the trivial representation
$r={\bf 1}$, the defining, fundamental representation $r={\bf N}$, and its
complex conjugate $r={\bf N}^*$ are needed. All sites are
projected onto the trivial representation except at most a few onto these
fundamental representations.
The line expectation values are expressed in terms sums of diagonal matrix
elements of projection operators.
\begin{equation}
 \langle L(i) L(j) \ldots \rangle = \frac{
Tr[e^{-H/T}P(i,{\bf N})P(j,{\bf N})\ldots\prod_{k\neq i,j,\ldots}P(k,{\bf 1})]}
 {Tr[ e^{-H/T} {\cal P}]}
\end{equation}

In this view, the physical variables that describe a state of the quantum
system
are the group elements on spatial links. The group elements on 4-direction
links are unphysical, auxiliary variables used in a technical way to enforce
the
Gauss law constraints.

\subsection{Global symmetry}

Now let $z$ stand for one of the $N$ elements of the center $Z(N)$ of $SU(N)$.
Consider the global transformation $\bar{V}(z) = \prod_i V(i,z)$.
In either basis, this is easily seen to act trivially on the physical
variables.
For example, with definite group elements $u(l)$ on spatial links $l$ bounded
by sites $i$ and $i'$,
\begin{equation}
   u(l) \rightarrow g^{-1}(i) u(l) g(i') \rightarrow z^{-1} u(l) z = u(l)  .
\end{equation}
Similarly, when the flux is definitely in representation $r(i)$
with indices $m$ and $n$,
the amplitude
for the group element on
link $l$ to have a value $u(l)$ is given by the representation matrix
$D_m^n(u(l),r(i))$. The relations
\begin{equation}
  D(z^{-1} u(l) z,r(i)) = D^{-1}(z,r(i)) D(u(l),r(i)) D(z,r(i)) = D(u(l),r(i))
\end{equation}
show again that the action of $\bar{V}(z)$ is trivial. In the
absence of a nontrivial, physical, global transformation, there can be no
physical
symmetry breaking at the phase transition. The confining and nonconfining
phases are not distinguished by different physical symmetry realizations.

Since the operator $\bar{V}(z)$ acts as 1, it can be inserted into any matrix
element without changing its value. However, because
\begin{equation}
   \bar{V}(z) V(g) =\bar{V}(z) \prod_i V(i,g(i)) = \prod_i V(i,zg(i))  ,
\end{equation}
there is the effect of changing each auxiliary
variable from $g(i)$ to $z g(i)$.
With fixed $g$'s, the weight for a flux configuration is
\begin{equation}
 \langle \theta |e^{-H/T} \prod_i V(i,g(i)) | \theta \rangle  .
\end{equation}
Under the global transformation $g \rightarrow z g$, it is unchanged:
\begin{eqnarray}
 \langle \theta |e^{-H/T} \prod_i V(i,g(i)) | \theta \rangle \rightarrow
 \langle \theta |e^{-H/T} \prod_i V(i,zg(i)) | \theta \rangle
 & = & \langle \theta |e^{-H/T} \bar{V}(z) \prod_i V(i,g(i)) | \theta \rangle\\
 & = & \langle \theta |e^{-H/T} \prod_i V(i,g(i)) | \theta \rangle . \label{e6}
\end{eqnarray}
In the path integral, this becomes the global
$Z(N)$ symmetry of the 4-direction link variables in the classical, Lagrangian
view. Thus, the ordering of the Wilson lines does not break any physical
symmetry.

Furthermore, a similar argument using $\bar{V}(z)$ gives
\begin{equation}
   0 = \int \!\! dg \, \chi^*_{{\bf N}}(g) \, \langle \phi|V(i,g)|\psi \rangle
\end{equation}
for any physical states $\phi$ and $\psi$. Thus, any calculation that gives
$\langle L \rangle \neq 0$ cannot be interpreted as
\begin{equation}
 Tr [ e^{-H/T} P(i,{\bf N}) \prod_{j \neq i} P(j,{\bf 1})] .
\end{equation}
To find the correct interpretation of $\langle L \rangle \neq 0$, the external
field is introduced again.

\subsection{Interpretations}

At this point, the relation of the global symmetry of the Wilson lines to the
physical, Hamiltonian formulation has been given.
As indicated in (\ref{e6}), when the auxiliary
variables $g(i)$ that enforce the Gauss law constraint and become the Wilson
lines are multiplied by an element of the center of the group, the weights for
the different physical configurations of flux are unchanged. Thus, the physical
configurations are insensitive to the phase of $\langle L \rangle$ in the
broken-symmetry region.
Each of the $N$ ordered, pure phases gives the same result for physical
quantities. In calculating physical quantities, one could just as well sum
over these pure phases and have $\langle L \rangle = 0$.

To treat this more carefully and find an interpretation for
$\langle L \rangle \neq 0$,
an external field $h$ is introduced again.
Let $L(i)$ be the line at site $i$ computed in the defining, fundamental
representation {\bf N}.
The logarithm of the Boltzmann weight gets the additional term
$h^* \sum_i L(i) + h \sum_i L^*(i)$. Now $g \rightarrow zg $
is equivalent to
$h \rightarrow z^*h $, so the free energy must be invariant under a
rotation of $h$ by an element of $Z(N)$. Conversely, modifying $h$ by such a
factor in the Boltzmann weight has no effect on the weights of the different
physical configurations.

In a finite volume with $h \rightarrow 0$, there is no singularity, and the
$Z(N)$ invariance guarantees $\langle L \rangle = 0$. In that case, there is no
contradiction to resolve.
The problem is in infinite volume where $\langle L \rangle \neq 0$ is possible.
An interpretation of $\langle L \rangle$ as the expectation value of a
projection operator or as $e^{-F/T}$ requires $\langle L \rangle \geq 0$.
Also, $Z(N)$ invariance requires $\langle L \rangle = 0$.
Thus, a negative or complex
value for $\langle L \rangle$ requires some explaining.

Of course, the problem can be avoided again by ignoring it and extracting the
desired information from the two-point function at infinite separation
$\langle L(0) L^*(\infty) \rangle$. But this does not do much to refine our
understanding or resolve the contradiction.

With an extra factor of $e^{h^* \sum_i L(i) + h \sum_i L^*(i)}$ inserted in the
Boltzmann weight, there is a finite density of sites projected onto nontrivial
representations. Again, the character expansion makes this precise.
At each site, the factor is
\begin{eqnarray}
  e^{h^* L(i) + h L^*(i)} & = & \sum_{r} a_r(h,h^*) \chi_r(g(i))  \\
 & = & a_{{\bf 1}} (1 + \sum_{r \neq {\bf 1}} \frac{a_r}{a_{{\bf 1}}} \chi_r)
\\
      & = & a_{{\bf 1}} (1 + \frac{a_{\bf N}}{a_{{\bf 1}}} \chi_{\bf N}
               + \frac{a_{{\bf N}^*}}{a_{{\bf 1}}} \chi_{{\bf N}^*}) + O(h^3)
\end{eqnarray}
with
\begin{equation}
 \frac{a_{\bf N}}{a_{{\bf 1}}} = h^*/N + O(h^2) \;\; \text{ and } \;\;
 \frac{a_{{\bf N}^*}}{a_{{\bf 1}}} = h/N + O(h^2)    .
\end{equation}
The density of sites in the fundamental representation is $O(h^*)$.

Thus, the expectation value of the line is computed in the presence of a finite
density of other sites in nontrivial representations. As $h \rightarrow 0$,
this density goes to zero. The small $h$ expansion starts as
\begin{equation}
 \langle L(0) \rangle_h = h \sum_i \langle L(0) L^*(i)\rangle_0 + \ldots
                                  \;\; .                         \label{e7}
\end{equation}

The large-volume limit of this function depends on the phase of $h$ so that as
$h \rightarrow 0$, the result is proportional to the nearest element of $Z(N)$.
For example,
\begin{equation}
 \langle L(0) \rangle_h = \frac{1}{\cal V} \frac{\partial}{\partial h^*} \ln Z
\end{equation}
with
\begin{equation}
 Z = \sum_{z \in Z(N)} e^{{\cal V} |\langle L \rangle|(h^* z + h z^*)}  .
\end{equation}

The large-volume limit depends on the phase of $h$. However, each matrix
element
that goes into that limit of (\ref{e7}) is computed in finite volume
with no external field
and is a global $Z(N)$ singlet. The density of the nontrivial sites goes
to zero, and those with $i \neq 0$ recede to infinity as $h \rightarrow 0$.
The contribution from each flux
configuration is unchanged by the replacement $h \rightarrow z h$.
Thus, the flux configurations and all physical properties are
independent of the phase of $\langle L \rangle$.
The conclusion is that
\begin{equation}
 \langle L(i) \rangle = z e^{-F/T}    .
\end{equation}
$F$ is the excess free energy associated with a source at $i$ in a background
of other sources as their density goes to zero. The phase $z \in Z(N)$ is
determined by the direction from which the external field approaches zero.

To confirm that there is no change in the realization of the gauge properties
of the theory in the nonconfining phase, one needs to check that even though
the global $Z(N)$ is broken, there is no effect on the local Gauss law
constraints. Since it has already been shown that global $\bar{V}(z)$ has no
physical effect, this is a pretty safe bet.

The constraint is supposed to project onto the trivial representation of
$SU(N)_i$. If that is true, then
\begin{equation}
  \langle V(i,g') \rangle = 1 \;\;  . \label{e8}
\end{equation}
The effect of this extra operator is to multiply the auxiliary
variable $g(i)$ by $g'$. However, that effect can be immediately removed by a
change of variable in the $g(i)$ integration. Thus, (\ref{e8})
is correct, and the
Gauss law is implemented. This works whether or not the $g$'s have ordered.

If you are nervous about that in the ordered phase, consider it more carefully
in the
presence of the external field. In that case, there is additional $g(i)$
dependence in the weight from $e^{h^* \sum_i L(i) + h \sum_i L^*(i)}$. Since
$L(i)$ is the trace of $g(i)$ in the defining representation, when the
integration variable is changed, there is a new factor
\begin{equation}
    e^{h^* (z-1) L(i) + h (z^*-1)L^*(i)}   \label{e9}
\end{equation}
associated with site $i$ only. This is a local change, so the global
phase is still determined by the field $h$ acting over the whole
volume. The factor (\ref{e9}) becomes 1 as $h \rightarrow 0$, the
expectation value of $V$ is 1, and the local Gauss law is correctly
implemented even in an ordered, pure phase.

To complete the parallel with the Ising flux model, the analogue of the Ising
spins view must be mentioned. For gauge theory, this view is known as
the effective theory for the Wilson lines\cite{s3}. Holding the
values of the lines fixed, all the other variables are integrated out. This
leaves a theory in three space dimensions with fields $L(i)$ on sites.
There is a global $Z(N)$ symmetry.

\subsection{Monte Carlo}

Monte Carlo calculations are ordinarily done in finite volume with no external
field. How should they be interpreted? If one were to run long enough, the
result would be $\langle L \rangle = 0$. However, in practice, in the
nonconfining phase and not too close to the phase transition, the
flips of $L$ from one pure phase to another occur
infrequently or not at all so that
$\langle L \rangle \neq 0$ for a run. This has no effect on the measurement
of physical variables. A series of configurations in which
the line does not flip corresponds to the $h \rightarrow 0$ after
${\cal V} \rightarrow \infty$ procedure. They must be interpreted according to
the discussion in the previous subsection. They cannot be given the simple
interpretation
\begin{equation}
 \langle L(i) \rangle \sim
              Tr [ e^{-H/T} P(i,{\bf N}) \prod_{j \neq i} P(j,{\bf 1})] .
\end{equation}

\section{Percolating flux}

I have asserted that the difference between the confining and the nonconfining
phases is not a difference in the realization of a physical symmetry but rather
the presence or absence of a network of percolating flux. Let me elaborate.

\subsection{Ising flux}

It is helpful to look at Ising flux first. Consider two widely
separated points $i$ and $j$ and the correlation function
$\langle s(i) s(j)\rangle$.
In the low-temperature phase, flux is sparse, and the character expansion is
convergent.
If the sites $i$ and $j$
are not connected by any flux, then the spin sums in the separate clusters that
contain $i$ and $j$ are independent, and the contribution to
$\langle s(i) s(j) \rangle$ is zero.
When the probability of arbitrarily large flux clusters
goes to zero as their size increases, the correlation function must vanish
as the separation becomes arbitrarily large. Thus, no percolating cluster
implies no ordering.

On the other hand, if there is ordering so that
\begin{equation}
 \lim_{|i-j| \rightarrow \infty} \langle s(i) s(j) \rangle
             = \langle s \rangle^2 > 0 ,
\end{equation}
then there must be a finite probability to find $i$ and $j$ in the same cluster
no matter how large $|i-j|$, i.e. there is a percolating cluster.
Spin ordering implies a percolating cluster.

Conversely, if there is a percolating cluster of finite density, then there is
a finite probability to find both $i$ and $j$ in it no matter how large
$|i-j|$.
It is necessary to consider the $|i-j|$ dependence of this probability.
I do not know how to address this rigorously, but consider the following
discussion:
In any finite volume,
the largest cluster hits a finite fraction of the sites. If that fraction
remains finite as the volume goes to infinity, then there is a percolating
cluster.
It seems clear that energy and entropy will conspire to give the percolating
cluster an approximately constant density. Large departures from constant
density have low entropy.
With energy and entropy conspiring to give constant density for the percolating
cluster, the disturbance caused by an odd-flux site will extend over a
fixed, finite distance. If $i$ and  $j$ are
separated by more than that, then further separation will have no effect.
Assume that the weight for a largest cluster is a function of its density.
Since there are largest clusters with the same density independent of
$|i-j|$ when it is large, the probability to find both $i$ and $j$ in the
largest cluster is also independent of $|i-j|$.

After the spin sums at sites other than $i$
and $j$ have been done, one is left with a factor proportional to $[s(i) s(j)]$
as the contribution from all the links in the largest cluster. The final sums
are
\begin{equation}
 \langle s(i) s(j) \rangle \sim
              \sum_{s(i)} \sum_{s(j)} s(i) [s(i) s(j)] s(j) > 0   .
\end{equation}
Finally, take the infinite-volume limit. This shows that
percolating flux implies spin ordering.
Note that the choice of a pure phase for the spins causes no difficulties
because the two-point function and the flux distributions are independent of
that choice.

In conclusion, the physical difference between the nonconfining and the
confining phases is the presence of percolating flux.

\subsection{Gauge flux}

After the discussion above, it is easy to apply the ideas to gauge flux.
The flux on links is more complicated since it is labeled by the irreducible
representations of $SU(N)$ rather than $Z(2)$. However, the argument above is
geometrical in its essentials and does not depend upon the detailed
nature of the flux. It is necessary that the auxiliary variable group
elements $g(i)$ and $g(j)$ be uncorrelated if $i$ and $j$ are in different
clusters. This is certainly true.
Define ``there is a percolating cluster" to
mean that percolating clusters make a
finite contribution to
$\langle L(i) L(j) \rangle$
when the separation is {\em finite}. Now, no matter how large the
separation, there is a finite probability to have both sites hit by the
percolating cluster in a flux configuration. If the disturbance in the cluster
caused by one nontrivial site has a finite range, then the
value of the contribution from the percolating clusters is independent of
$|i-j|$ for large $|i-j|$.
Thus, the reasoning from the Ising flux case can be used again. The conclusion
is that the physical difference between
the nonconfining and confining phases is the presence of percolating flux.

\section{Summary}

The usual relation $\langle L \rangle = e^{-F/T}$ cannot be correct for
nonzero, nonpositive values of $\langle L \rangle$.
A careful approach to the $Z(N)$ symmetry and to the
infinite-volume limit resolves the problem. In terms of the physical
variables of quantum, finite-temperature, Hamiltonian gauge fields,
there is no $Z(N)$ symmetry to be broken. The physical system is the same in
each of the $N$ pure phases that can be chosen when $\langle L \rangle \neq 0$.
The confining and nonconfining phases are distinguished by the presence of
percolating flux in the nonconfining phase. To
treat the infinite-volume, $\langle L \rangle \neq 0$ pure phases, an external
field is introduced
in the usual manner. This leads to the conclusion that
$\langle L(i) \rangle = z e^{-F/T}$.
The element $z \in Z(N)$ is determined by the
external field. $F$ is the zero-density limit of the excess free energy of a
fundamental source of flux at site $i$ in a background of other sources.

\acknowledgments

I thank R. Narayanan and P. Vranas for helpful comments on the manuscript.
This research was supported by the United States Department of Energy.

\end{document}